
\bigskip\bigskip

INFN-NA-IV-93/35~~~~~~~~~~~~~~~~~~~~~~~~~~~~~~~~~~~~~~~~~~~~~~DSF-T-93/35

\vspace{4cm}

\vbox{\vspace{38mm}}
\begin{center}
{\LARGE \bf Photon distribution for one-mode mixed light with generic gaussian
wigner function\\[2mm]}\\[5mm]
\end{center}
\begin{center}
V.V.Dodonov,O.V.Man'ko\\{\it Lebedev Physics Institute,Moscow,
Russia}\\[3mm] and V.I.Man'ko\\{\it Dipartimento di Scienze Fisiche
Universita di Napoli "Federico II"and I.N.F.N.,Sez.di Napoli
\end{center}

\newpage

\begin{abstract}

For one mode light described by the Wigner function of generic Gaussian
form the photon distribution function is obtained explicitly and expressed
in terms of Hermite polynomial of two variables.The mean values and dispersions
of photon numbers are obtained for this generic mixed state.Generating
function for photon distribution is discussed.Known partial
cases of thermal state,correlated state,squeezed state, and coherent
state are considered.The connection of Schrodinger uncertainty relation
for quadratures with photon distribution is demonstrated explicitly.

\end{abstract}

\newpage

\section{Introduction}

The problem of finding the photon distribution function for the one-mode
electromagnetic field described by generic
Gaussian Wigner function corresponding to the most general mixed gaussian
state
has been concidered in Ref.[1-5],and some special cases were investigated
also in Ref.[6-8].The expressions
for the photon distribution function were obtained as a rule
in the forms of
series,and the simplest form found in the Ref.[4] was the finite series of
the products of Laguerre polynomials.On the other hand,the
Q-function of the density operator, which corresponds to the exponential of
generic quadratic form in the photon creation and annihilation
operators,has been calculated explicitly as far as in Ref.[9].In that
paper the diagonal elements of the
density operator in the number state basis were expressed
in terms of Hermite polynomials of two variables for one-mode field.
Similar results, but in much more special cases, were given in Ref.[6,10].
Q-function is the matrix element of the density operator in the
coherent state basis,it is the generating function for the matrix
elements of the density operator in the number state basis.Consequently,
using
Q-function one can obtain the generating function for the photon
distribution function,and this was done in Ref.[4],[9].

The aim of our work is to show that the photon distribution function
of the one-mode field state described by the generic Gaussian Wigner
function may be calculated explicitly in terms of
Hermite polynomials of two variables with equal indices.So,the combersome
series
found in Ref.[1-4] may be simplified
significantly.
In particular the finite series of products of two Laguerre polynomials
found in Ref.[4] is simply proportional to the Hermite polynomial of two
variables with equal indices.

We show that all the known limiting cases of the photon probability
functions discussed in Ref.[1-8] may be obtained as partial cases of the
general
expression in terms of the Hermite polynomial of two variables.These
limiting cases of the one-mode field are:the coherent light[11],
the squeezed light[12-15],the correlated coherent light[16],
the thermal equilibrium state (black body radiation),
the squeezed thermal state[5-8].

We will clarify the physical meaning of the parameters which the
photon distribution function depends on,using the Wigner representation for the
field density operator since in this representation it is obvious that
the generic case of Gaussian Wigner function contains five real parameters.
Three
of these parameters are the variances and the covariance
of photon quadrature components.Two others are the means of the quadratures.
Our goal is to express the photon distribution function in terms of these
five parameters explicitly.

\section{Gaussian Wigner function}
The mixed squeezed state of the light with density operator $\hat \rho $ is
described by Wigner function $W(p,q)$ of the generic Gaussian form
which contains five real parameters.Two parameters are mean
values of momentum $<p>$ and position $<q>$ and other three parameters are
matrix elements of the real dispersion matrix $m$ with matrix elements
\begin {eqnarray}
m_{11}=\sigma _{pp},\\
m_{12}=\sigma _{pq},\\
m_{22}=\sigma _{qq}.
\end {eqnarray}
Below we will use the invariant parameters
\beq
T=Tr~{m}=\sigma _{pp}+\sigma _{qq}
\eeq
and
\beq
d=\det m=\sigma _{pp}\sigma _{qq}-\sigma _{pq}^2.
\eeq
The generic gaussian Wigner function has the form (see,for example,[17])
\beq
W(p,q)=d^{-\frac{1}{2}}\exp [-(2d)^{-1}[\sigma _{qq}(p-<p>)^{2}+\sigma _{pp}(q-
<q>)^{2}-2\sigma _{pq}(p-<p>)(q-<q>)]]
\eeq
The parameters $<p>$ and $<q>$ are given by the formulae
\begin {eqnarray}
<p>=Tr~ \hat \rho \hat p ,\\
<q>=Tr~ \hat \rho {\hat q},
\end {eqnarray}
where the operators $\hat p$ and $\hat q$ are the guadrature components of
photon creation $a\dag $ and the annihilation $a$ operators
\begin {eqnarray}
\hat p=\frac{a-a\dag}{i\sqrt 2},\\
\hat q=\frac{a+a\dag}{\sqrt 2}.
\end {eqnarray}
The matrix elements of the real symmetric dispersion matrix $m$ are defined
as follows
\begin {eqnarray}
\sigma _{pp}=Tr~ {\hat \rho}{\hat p}^2-<p>^2,\\
\sigma _{qq}=Tr~ {\hat \rho}{\hat q}^2-<q>^2,\\
\sigma _{pq}=\frac{1}{2} Tr~ {\hat \rho}({\hat p\hat q}
+{\hat q\hat p})-<p><q>.
\end {eqnarray}
Due to the physical meaning of the dispersions the parameters $\sigma _{pp}$
and $\sigma _{qq}$ must be nonnegative numbers,so the invariant parameter
$T$ (4) is a positive number.Also the determinant $d$ (5) of the dispersion
matrix must be positive.The function (6) if one does not precise that it
is Wigner function of a quantum system may be also concidered formally as
classical Gaussian distribution function.Then the above constrains of the
parameters domain take place for such formal distribution function.And
in classical version of the function (6) there is no others constrains.But
for quantum Wigner function the uncertainty relation provides additional
restriction and the domain of permitted values of the matrix elements of
the dispersion matrix is smaller for quantum Wigner function than for
classical Gaussian distribution function in phase space of a particle.
Below we will discuss this difference and connect it with photon
number distribution function.

\section{Photon distribution function}

To obtain the photon distribution function we have to calculate the probability
$P _{n}$ to have $n$ photons in the state with the density operator $\hat
\rho$.
This probability is given by the formula
\beq
P_ {n}=Tr~ {\hat \rho}|n><n|,\\n=0,1,2,...,
\eeq
where the number states $|n>$ are the eigenstates of the number operator
$a\dag a$
\beq
a\dag a|n>=n|n>.
\eeq
The function $P_ n$ may be obtained if one calculates the generating function
for the matrix elements $\rho _{mn}$ of the density operator $\hat \rho$ in
the Fock basis.This generating function is the matrix element of the density
operator in the coherent state basis
\beq
<\beta|\hat \rho|\alpha>=\exp (\frac {-|\alpha|^2}{2}-\frac {|\beta|^2}{2})
 \sum_{m,n=0}^{\infty }{\frac {\beta ^{*m}\alpha ^n}{(m!n!)^{\frac {1}{2}}}
\rho _{mn}}
\eeq
and
\beq
P_ n=\rho _{nn}.
\eeq
The function $<\alpha |\hat \rho |\alpha >$ is the Q-function of
the system with the density operator $\hat \rho $.
The coherent state $|\alpha>$ is the normalized eigenstate of the
annihilation operator
\beq
a|\alpha>=\alpha|\alpha>.
\eeq
In terms of Wigner function the density operator in the coherent state
representation has the form of overlap integral [17]
\beq
<\beta |\hat \rho|\alpha>=\frac {1}{2\pi }\int W(p,q)W_{\alpha \beta }
(p,q)dpdq,
\eeq
where the function $W_{\alpha \beta }(p,q)$ is the Wigner function of the
operator $|\alpha ><\beta |$.It has the form [17]
\beq
W_{\alpha \beta }(p,q)=2\exp [-\frac {|\alpha |^2}{2}-\frac {|\beta |^2}{2}-
\alpha {\beta }^{*}-p^{2}-q^{2}+\sqrt 2\alpha (q-ip)+\sqrt 2{\beta }^{*}
(q+ip)].
\eeq
Since the integral given by Eq.(19) is the Gaussian one it may be easily
calculated.So,we have $$<\beta |\hat \rho |\alpha >=
P_{0}\exp (-\frac{|\alpha |^2}{2}-\frac{|\beta
|^2}{2})\exp [-\frac{1}{2}({\beta }^{*2}R_{11}+2\alpha {\beta }^{*}R_{12}
+\alpha ^{2}R_{22})+$$
\beq
+{\beta }^{*}y_{1}R_{11}+\alpha y_{2}R_{22}+R_{12}({\beta }^{*}y_{2}
+\alpha y_{1})].
\eeq
where the symmetric matrix $R$ has the matrix elements expressed in terms
of the dispersion matrix $m$ as follows
\begin {eqnarray}
R_{11}=(T+2d+\frac{1}{2})^{-1}(\sigma _{pp}-\sigma _{qq}-2i\sigma _{pq})
=R_{22}^*,\\
R_{12}=(T+2d+\frac{1}{2})^{-1}(\frac{1}{2}-2d).
\end {eqnarray}
This function coincides with the generating function for Hermite polynomials
of two variables[17],
and arguments of the Hermite polynomials are of the form
\beq
y_1={y_2}^*=-(T-2d-\frac{1}{2})^{-1}[(T-1)<z^*>+(\sigma _{pp}-\sigma _{qq}
+2i\sigma _{pq})<z>].
\eeq
The complex parameter $<z>$ is given by the relation
\beq
<z>=2^{-\frac{1}{2}}(<q>+i<p>).
\eeq
The probability to have no photon is given by the formula
\beq
P_0=(d+\frac{1}{2}T+\frac{1}{4})^{-\frac{1}{2}}\exp [(-<p>^{2}(\sigma _{qq}+
\frac{1}{2})-<q>^{2}(\sigma _{pp}+\frac{1}{2})+2\sigma _{pq}<p><q>)(\frac
{1}{2}+T+2d)^{-1}].
\eeq
Comparing the formula (21) with the generating function for Hermite
polynomials of two variables [18]
\beq
\exp [-\frac{1}{2}(R_{11}\beta ^{*2}+R_{22}\alpha ^{2}+2R_{12}\alpha \beta ^
{*})+R_{11}\beta ^{*}y_{1}+R_{22}\alpha y_{2}+R_{12}(\alpha y_{1}+\beta ^
{*}y_{2})]=\sum_{m,n=0}^{\infty }{\frac{H_{mn}^{\{R\}} (y_{1},y_{2})}
{n!m!}\alpha ^{n}\beta ^{*m}},
\eeq
we obtain for the photon distribution function $P_{n}$ the expression
\beq
P_{n}=P_{0}\frac{H_{nn}^{\{R\}}(y_{1},y_{2})}{n!}.
\eeq
Here the matrix $R$ determining the Hermite polynomial is given by the
formulae (22),(23) and two arguments of Hermite polynomial are given by the
expression (24).The expression (28) is the partial case of the matrix
elements of the density operator in Fock states basis obtained in [9]
by canonical transform method.Now we will derive the relation
of these two approaches based on Wigner function and on the canonical
transformation method.

\section{Canonical transform of thermal density operator}

We will use the canonical transformation of the photon creation and
annihilation operators determined by the unitary operator $K$
\beq
KaK\dag =ua+v{a\dag}+\delta ,
\eeq
where the complex numbers $u$,$v$ and $\delta $ are the parameters of
this transformation satisfying the condition
\beq
|u|^{2}-|v|^{2}=1,
\eeq
which preserves the boson commutation relations of the operators $a$ and
$a\dag $.The density operator $\hat \rho $ corresponding to the Wigner
function (6) may be obtained from the density operator $\hat \rho_{\beta}$
of the termal equilibrium state [9]
\beq
\hat \rho _{\beta }=2\sinh \frac{\beta }{2}\exp [-\beta (a\dag a+\frac{1}{2})].
\eeq
Here the parameter $\beta $ describes Planck distribution corresponding
to the mean photon number in thermal state with the temperature $\beta ^{-1}$
\beq
\bar n =\frac{1}{e^{\beta }-1}
\eeq
and with the mean energy in this state
\beq
E=\bar n +\frac{1}{2}.
\eeq
The connection of the density operator $\hat \rho $ with the operator
$\hat \rho _{\beta }$ is given by the formula
\beq
\hat \rho =K\hat \rho _{\beta} K\dag =2\sinh \frac{\beta }{2}\exp [-\beta (
u^{*}a\dag +v^{*}a+\delta ^{*})(ua+va\dag +\delta )+\frac{1}{2}],
\eeq
if one takes the following expressions for the parameters determining
the Wigner function (6) in terms of the parameters of the linear canonical
transformation (29)
\begin {eqnarray}
\sigma _{pp}=E(1+2|v|^{2} +2 Re~ {uv^{*}}),\\
\sigma _{qq}=E(1+2|v|^{2} -2 Re~ {uv^{*}}),\\
\sigma _{pq}=2E Im~ {uv^{*}},\\
<z>=-u^{*}\delta +v\delta ^{*}.
\end {eqnarray}
To describe the inverse transform we introduce the parameters:
\begin {eqnarray}
E=\sqrt d,\\
\sinh r=|v|=\frac {1}{2}\sqrt {Td^{-\frac{1}{2}}-2},\\
\cosh r=|u|=\frac {1}{2}\sqrt {Td^{-\frac{1}{2}}+2},\\
\sin (\phi _{u} - \phi _{v})=\frac {2\sigma _{pq}}{\sqrt {T^{2}-4d}},\\
\delta =-u<z>-v<z^{*}>.
\end {eqnarray}
Here  $\phi _{u}$ and $\phi _{v}$ are phases of complex numbers $u$ and
$v$.If we have the density matrix (34) the parameters of Wigner function
(6) corresponding to the density matrix are given by the formulae (35)-(38).
They depend on the phase difference $\phi _{u}-\phi _{v}$ only.Due to that
we will take
\begin {eqnarray}
u=|u|=\cosh r,\\
v=-\sinh r e^{i\theta },
\end {eqnarray}
where
\beq
\theta =\phi _{v} -\pi .
\eeq
This choice corresponds to the representation of the operator $K$ in the form
of the product of the squeezing operator
\beq
S=\exp [\frac{1}{2}r(e^{i\theta }a\dag ^{2}-e^{-i\theta }a^{2})]
\eeq
and the displacement operator
\beq
D(-\delta )=\exp (-\delta a\dag +\delta ^{*}a).
\eeq
So,we have for the density operator the factorised form
\beq
\hat \rho =SD(-\delta )\hat \rho _{\beta}D(\delta )S\dag,
\eeq
or changing the order of the operators
\beq
\hat \rho =D(<z>)S\hat \rho _{\beta}S\dag D\dag (<z>).
\eeq
The matrix elements of this operator in Fock basis were obtained in [9]
and they are expressed in terms of Hermite polynomials of two
variables
\beq
<m|\hat \rho |n>=P_{0}(m!n!)^{-\frac{1}{2}} H_{mn}^{\{R\}} (y_{1},
y_{1}^{*}),
\eeq
where the symmetric matrix $R$,the argument of the Hermite polynomial
$y_{1},y_{2}$
and the factor $P_{0}=<0|\hat \rho |0>$ are expressed in terms of Wigner
function parameters due to the formulae (22) -(26).The product form of
the density operator $\hat \rho $ (50) was derived in Ref.[4].

\section{Generating function for some Hermite polynomials}

Let us derive new formulae for Hermite polinomials $H_{nn}^{\{R\}} (y_{1},
y_{2})$ which are defined by means of a symmetric matrix
\beq
R = \left( \begin{array}{clcr}
    R_{11} & R_{12}\\
    R_{21} & R_{22}\end{array} \right).
\eeq
The generating function for Hermite polynomials $H_{nm}^{\{R\}} (y_{1},
y_{2})$ is given in the form [18]
\beq
\exp [-\frac{1}{2} {\bf a} R{\bf a}+{\bf a}R{\bf y}] = \sum_{n,m=0}^{\infty}
\frac{a_{1}^{n} a_{2}^{m}}{n!m!} H_{nm}^{\{R\}} (\bf y ).
\eeq
Here ${\bf a}= (a_{1},a_{2})$ and the numbers $a_{1}$ and $a_{2}$ are
arbitrary complex numbers:
\beq
{\bf a} R{\bf a} = \sum_{i,k=0}^{2} a_{i}R_{ik}a_{k}
\eeq
and
\beq
{\bf a} R{\bf y} = \sum_{i,k=0}^{2} a_{i}R_{ik}y_{k}.
\eeq
Let us take $\bf a$ in the expression (53) to have components
\beq
a_{1}=\beta ^{*},~~~a_{2}=\lambda \beta ,
\eeq
where $\lambda $ is the new complex parameter.Then multiplying
both sides of the equation (53) by the factor $\exp (-\beta \beta ^{*})$
and calculating the gaussian integral
\beq
G(\lambda )=\frac{i}{2\pi }\int \exp (-\frac{1}{2}{\bf {a}}R{\bf {a}}+{\bf {a}}
{\bf {y}}-\beta \beta ^{*})d\beta d\beta ^{*}=\sum_{n,m=0}^{\infty} \lambda ^
{m}\frac{H_{nm}^{\{R\}}(R^{-1}{\bf y })}{n!m!} \frac{i}{2\pi }
\int \beta ^{*n} \beta ^{m} e^{-\beta\beta ^{*}} d\beta d\beta ^{*},
\eeq
we obtain the relation for the function
\beq
G(\lambda )=\sum_{n=0}^{\infty}\frac{H_{nn}^{\{R\}}(R^{-1}{\bf y})}{n!}
\lambda ^{n},
\eeq
since
\beq
\frac{i}{2\pi }\int \beta ^{*n}\beta ^{m}e^{-\beta \beta ^{*}}d\beta d\beta
^{*}
=n!\delta _{nm}.
\eeq
This relation means that $G(\lambda )$ is the generating function for the
Hermite
polynomials of two variables with equal indices.The Gaussian integral for the
generating function $G(\lambda )$ may be easily calculated and we have
\beq
G(\lambda )=i[\det(\lambda R+\sigma _{x})]^{-\frac{1}{2}}\exp[\frac{1}{2}{\bf
y}
(R+\lambda ^{-1}\sigma _{x})^{-1}{\bf y}],
\eeq
where the matrix $\sigma _{x}$ is the Pauli matrix
\beq
\sigma _{x}= \left( \begin{array}{clcr}
0&1\\
1&0\end{array}\right).
\eeq
Comparing the relation (58) with the photon distribution function (28) we
see that the function $G(\lambda )$ is in fact the generating function
for the photon distribution function.In other notations this function has
been derived in Ref.[9].

\section{Sum rules for Hermite polynomials}

So, we have the sum rule
\beq
\sum_{n=0}^{\infty}\lambda ^{n}\frac{H_{nn}^{\{R\}}(R^{-1}{\bf y})}{n!}=i[\det(
\lambda R+\sigma _{x})]^{-\frac{1}{2}}\exp[\frac{1}{2}{\bf y}(R+\lambda ^{-1}
\sigma _{x})^{-1}{\bf y}].
\eeq
This sum rule may be generalized for $2N$-dimensional case by the same method.
We have the sum rule
\beq
\sum_{n_{1}=0}^{\infty}\ldots\sum_{n_{N}=0}^{\infty}\frac{\lambda _{1}^{n_{1}}}
{n_{1}!}\frac{\lambda _{2}^{n_{2}}}{n_{2}!}\ldots\frac{\lambda _{N}^{n_{N}}}
{n_{N}!} H_{n_{1}n_{2}\ldots n_{N}n_{1}n_{2}\ldots n_{N}}^{\{R\}}(R^{-1}{\bf y}
) =i^{N}[\det(\lambda R+\Sigma _{x})]^{-\frac{1}{2}}\exp[\frac{1}{2}{\bf y}
(R+\lambda ^{-1}\Sigma _{x})^{-1}{\bf y}],
\eeq
where ${\bf {y}}=(y_{1},y_{2},...y_{2N})$,the $2N$x$2N$ matrix $\Sigma _{x}$ is
the $2N$-dimensional analog of the Pauli matrix $\sigma _{x}$
\beq
\Sigma _{x} = \left( \begin{array}{clcr}
\bf 0 &\bf 1\\
\bf 1 &\bf 0\end{array}\right),
\eeq
and {\bf {1}} is $N$-dimensional unity matrix and the  diagonal $2N$x$2N$-
matrix $\lambda $ has the form
\beq
\lambda = \left( \begin{array}{clcr}
\bar \lambda&0\\
0&\bar \lambda \end{array}\right),
\eeq
where
\beq
\bar \lambda = \left( \begin{array}{clcr}
\lambda _{1}&\ldots&0\\
\vdots&\ddots&\vdots\\
0&\ldots&\lambda _{N}\end{array}\right).
\eeq
For the Hermite polynomials of two variables $H_{nn}^{\{R\}}(\bf y)$
there exists the sum rule [9],[17]
\beq
H_{nn}^{\{R\}}(y_{1},y_{2})=(n!)^{2} (\frac{R_{11}R_{22}}{4}
)^{\frac{n}{2}} \sum_{k=o}^{n} (-\frac{2R_{12}}{\sqrt {R_{11}R_{22}}}
)^{k}{(n-k)!}^{-2}{(k!)^{-1}{H_{n-k}(\frac{z_{1}}{\sqrt {2R_{11}}})}
{H_{n-k}(\frac{z_{2}}{\sqrt {2R_{22}}})},
\eeq
where $$z_{1}=R_{11}y_{1}+R_{12}y_{2},~~~z_{2}=R_{12}y_{1}+R_{22}y_{2}.$$
So,these polynomials are expressed in terms of usual Hermite polinomials
of one variable.We can obtain analogous formula in which the polynomials
$H_{nn}(y_{1},y_{2})$ are expressed in terms of Laguerre polynomials.
To do this we rewrite the right hand side of the formula (60) in the form
\beq
i[\det(\lambda R+\sigma _{x})]^{-\frac{1}{2}}\exp[\frac{1}{2}{\bf y}(R+
\lambda ^{-1}\sigma _{x})^{-1}{\bf y}]=[(1-\frac{\lambda }{\lambda _{1}})
(1-\frac{\lambda }{\lambda _{2}})]^{-\frac{1}{2}}\exp[\frac{\lambda x_{1}}
{\lambda -\lambda _{1}}+\frac{\lambda x_{2}}{\lambda -\lambda _{2}}],
\eeq
where
\begin {eqnarray}
\lambda _{1}=(\sqrt {R_{11}R_{22}}-R_{12})^{-1},\\
\lambda _{2}=-(\sqrt {R_{11}R_{22}}+R_{12})^{-1}
\end {eqnarray}
and
\begin {eqnarray}
x_{1}=\frac{1}{4}(R_{12}-\sqrt {R_{11}R_{22}})(2y_{1}y_{2}-\sqrt {\frac{R_{11}}
{R_{22}}} y_{1}^{2}-\sqrt {\frac{R_{22}}{R_{11}}} y_{2}^{2}),\\
x_{2}=\frac{1}{4}(R_{12}+\sqrt {R_{11}R_{22}})(2y_{1}y_{2}+\sqrt {\frac{R_{11}}
{R_{22}}} y_{1}^{2}+\sqrt {\frac{R_{22}}{R_{11}}} y_{2}^{2}).
\end {eqnarray}
Using the known relation [18] for Laguerre polynomials
\beq
\frac {1}{\sqrt {1-z}}\exp{\frac{xz}{z-1}} =\sum_{n=0}^{\infty }z^{n}L_{n}^{-
\frac{1}{2}} (x),
\eeq
we see that the generating function $G(\lambda )$ may be represented in the
form
\beq
G(\lambda ) = \sum_{s=0}^{\infty }(\frac{\lambda }{\lambda _{1}})^{s} L_{s}^{-
\frac{1}{2}} (x_{1}) \sum_{k=o}^{\infty }(\frac{\lambda }{\lambda _{2}})^{k}
L_{k}^{-\frac{1}{2}}(x_{2}).
\eeq
Thus we obtained the formula
\beq
H_{nn}^{\{R\}}(R^{-1}{\bf y}) = (-1)^{n} n!\sum_{s=0}^{n}(R_{12}
-{\sqrt {R_{11}R_{22}}})^{s}(R_{12}+{\sqrt {R_{11}R_{22}}})^{n-s}
L_{s}^{-\frac{1}{2}}(x_{1}) L_{n-s}^{-\frac{1}{2}}(x_{2}),
\eeq
where $x_{1}$ and $x_{2}$ are given by the formula (71),(72).
Using known relation[9],[17]
\beq
H_{nn}^{\{R\}}({\bf 0})=(-1)^{n}n!(\beta ^{2}-1)^{\frac{n}{2}}P_{n}
(\frac{\beta }{\sqrt {\beta ^{2}-1}}),
\eeq
which takes place for~~$R_{11}=R_{22}=1,~~R_{12}=\beta $,~~we have the sum rule
\beq
(\beta ^{2}-1)^{\frac{n}{2}}\sum_{s=0}^{n}(\beta +1)^{-s}(\beta -1)^{n-s}
L_{s}^{-\frac{1}{2}}(0)L_{n-s}^{-\frac{1}{2}}(0)=P_{n}(\frac{\beta }
{\sqrt {\beta ^{2}-1}}),
\eeq
where $P_{n}(x)$ is Legendre polynomial.
For the matrix $R$ proportional to the unity matrix $R=R_{11}{\bf 1}$ and
argument of Hermite polynomial of two variables equal to zero
$({\bf y}=0)$ one can obtain the equality
\beq
H_{2n2m}^{\{R\}}(0,0)=(-1)^{n+m}2^{-n-m}{\frac{2n!2m!}{n!m!}}R_{11}^{n+m}
\eeq
and
\beq
H_{2n,2k+1}^{\{R\}}(0,0)=0.
\eeq
If for the same matrix $R$ the argument of Hermite polynomial is equal to
${\bf y}$ we have the relation
\beq
H_{nm}^{\{R\}}(y_{1},y_{2})=(\frac{R_{11}}{2})^{\frac{n+m}{2}} H_{n}(
\sqrt {\frac{R_{11}}{2}}y_{1}) H_{m}(\sqrt {\frac{R_{11}}{2}}y_{2}).
\eeq
In the limit $R{\rightarrow}0$ and the argument ${\bf y}$ is such that
\beq
R{\bf y}={\bf z}
\eeq
where the vector ${\bf z}$ has the fixed components $(z_{1},z_{2})$ we
have the relation
\beq
H_{nn}^{\{0\}}({\bf y})=(z_{1}z_{2})^{n}.
\eeq
If the matrix $R$ has diagonal form
\beq
R=\left( \begin{array}{clcr}
r & 0\\
0 & r^{*}\end{array} \right)
\eeq
and the vector ${\bf y}=(y,y^{*})$ where the numbers $r,r^{*},y,y^{*}$
are any complex numbers we have the relation
\beq
H_{nm}^{\{R\}}(y,y^{*})=(\frac{r}{2})^{\frac{n}{2}}(\frac{r^{*}}{2})^{
\frac{m}{2}}H_{n}({\sqrt \frac{r}{2}}y)H_{m}({\sqrt \frac{r^{*}}{2}}y^{*}).
\eeq
Since $H_{2n+1}(0)=0$ we obtain for diagonal matrix $R$ the equality
\beq
H_{2n+1,m}^{\{R\}}(0,0)=0.
\eeq
If we have two matrices
\beq
R=\sigma _{x},~~~~R_{1}=t^{2}\sigma _{x},
\eeq
where $\sigma _{x}$ is the Pauli matrix one can show that
\beq
H_{nm}^{\{R_{1}\}}(y_{1},y_{2})=t^{n+m}H_{nm}^{\{R\}}(ty_{1},ty_{2})
\eeq
and
\beq
H_{nn}^{\{R_{1}\}}(y_{1},y_{2})=t^{2n}H_{nn}^{\{\sigma _{x}\}}(ty_{1},
ty_{2}).
\eeq
On the other hand there exists the relation [9],[17]
\beq
H_{nn}^{\{\sigma _{x}\}}(z_{1},z_{2})=(-1)^{n}n!L_{n}(z_{1}z_{2}),
\eeq
where $L_{n}(z)$ is the Lagerre polynomial.It means that
\beq
H_{nn}^{\{R_{1}\}}(y_{1},y_{2})=(-1)^{n}t^{2n}n!L_{n}(t^{2}y_{1}y_{2}).
\eeq
For ${\bf y}=0$ we have
\beq
H_{nn}^{\{R_{1}\}}(0,0)=(-1)^{n}t^{2n}n!.
\eeq
Using the generating function $G(\lambda )$ (58) one can find the sum rule
\beq
\sum _{n=0}^{\infty }\frac{nH_{nn}^{\{R\}}({\bf y})}{n!}}=\{\frac{1}{2}
{\bf y}R(\sigma _{x}R+1)^{-2}\sigma _{x}R{\bf y}-\det (R+\sigma _{+})(R+
\sigma _{x})^{-1}\}\sum_{n=0}^{\infty }{\frac{H_{nn}^{\{R\}}{(\bf y)}}{n!}}
\eeq
where $\sigma _{x}$ is the Pauli matrix and the matrix $\sigma _{+}$ has
the form
\beq
\sigma _{+}=\left( \begin{array}{clcr}
0&1\\
0&0\end{array} \right).
\eeq
Let us introduce "higher momenta" $<n^{k}>$ due to relation
\beq
\sum_{n=0}^{\infty } \frac{n^{k}H_{nn}^{\{R\}}({\bf y})}{n!}=<n^{k}> \sum_{n=0}
^{\infty } \frac{H_{nn}^{\{R\}}({\bf y})}{n!}
\eeq
and the "dispersion"
\beq
\sigma _{n}=<n^{2}>-<n>^{2}.
\eeq
Then for any Hermite polynomials $H_{nn}^{\{R\}}({\bf y})$ we have sum rules
\beq
<n>=\frac{1}{2} {\bf y}R\mu ^{2}\sigma _{x}R{\bf y}-\nu
\eeq
and
\beq
\sigma _{n}=\frac{1}{2}{\bf y}R[2\mu ^{3}-\mu ^{2}]\sigma _{x}R
{\bf y}+2\nu ^{2}-\nu -\det (\frac {R}{R+\sigma _{x}})
\eeq
where the matrix $\mu $ is defined by the relation
\beq
\mu =(\sigma _{x}R+1)^{-1}
\eeq
and the number $\nu $ is given by the formula
\beq
\nu =\det (\frac{R+\sigma _{+}}{R+\sigma _{x}}).
\eeq
The matrix $\sigma _{x}$ is the Pauli matrix and the matrix $\sigma _{+}$ is
given by formula (93).

\section{Photon distribution for coherent light}

Let us consider known partial cases of photon distribution functions in
terms of general expression (28).For coherent state [11] the
dispersion matrix is
\beq
m_{c}=\frac{1}{2}{\bf 1}
\eeq
and the matrix $R$ determining the Hermite polynomials is equal to zero.
According to the formula (26) for the coherent state the probability to have
no photons is
\beq
P_{0}=\exp[-(\frac{<p>^{2}+<q>^{2}}{2})].
\eeq
In this case the formula (82) gives for the Hermite polynomial $H_{nn}^{\{R\}}
({\bf y})$ the limit expression
\beq
H_{nn}^{\{0\}}({\bf y})=(\frac{<p>^{2}+<q>^{2}}{2})^{n},
\eeq
corresponding to relation (25).Combining the above formulae for the photon
distribution function in coherent state $|\alpha ><\alpha |$ with the
Wigner function $W_{\alpha \alpha }$ given by the formula (20) we obtain
from the general expression (28) the usual Poisson distribution
\beq
P_{n}=\frac{1}{n!}\exp[-\frac{<p>^{2}+<q>^{2}}{2}](\frac{<p>^{2}+<q>^{2}}
{2})^{n}.
\eeq
Due to the formulae (98),(99) the matrix $\mu $ is the identity matrix,
the number $\nu $ is equal to zero and from the formulae (96),(97) we
have $<n>=\sigma _{n}$.

\section{Photon distribution for squeezed light}

For the squeezed and correlated [16],[17] state the dispersion matrix
is
\beq
m_{s}=\frac{1}{2} \left( \begin{array}{clcr}
{\cosh 2r+\cos \theta \sinh 2r} & {\sinh 2r\sin \theta }\\
{\sinh 2r\sin \theta } & {\cosh 2r-\cos \theta\sinh 2r}\end{array} \right).
\eeq
Here
\beq
\cosh 2r=(\sigma _{pp}+\sigma _{qq}),~~~~~\sin \theta =\frac{2\sigma _{pq}}
{\sqrt {(\sigma _{pp}+\sigma _{qq})^{2}-1}}.
\eeq
The corresponding matrix $R_{s}$ is of the form
\beq
R_{s}=\tanh r \left( \begin{array}{clcr}
e^{-i\theta }&0 \\
0&e^{i\theta }\end{array} \right).
\eeq
The parameters $r$ and $\theta $ determine the dispersions of quadrature
components and their correlation coefficient
\beq
k=\frac{\sigma _{pq}}{\sqrt{\sigma _{pp}\sigma _{qq}}}=\frac{\sinh 2r\sin
\theta }{\sqrt {\cosh ^{2}2r-\cos^{2}\theta \sinh ^{2}2r}}.
\eeq
For $\theta =0$ there is no correlation of the quadrature components $(k=0)$.
In this case to apply the general formula for the photon distribution (28)
we calculate the argument of Hermite polynomial using the expressions for trace
and determinants of the dispersion matrix $m_{s}$.So we have
\beq
T=\cosh 2r ,~~~d=\frac{1}{4}
\eeq
and
\beq
y_{1}=y_{2}^{*}=-\frac{<q>-i<p>}{\sqrt 2} -e^{i\theta }{\coth r}
(\frac{<q>+i<p>}{\sqrt 2}).
\eeq
Due to formula (26) for the probability to have no photon we obtain for
squeezed state the expression
\beq
P_{0}=\frac{1}{\cosh r}\exp \{-(\frac{<p>^{2}+<q>^{2}}{2})+\frac{\tanh r}{2}
[(<p>^{2}-<q>^{2})\cos \theta +2<p><q>\sin \theta ]\}.
\eeq
In this case the matrix $R_{s}$ determining the Hermite polynomial of two
variables is the diagonal one.Using general expression (28) we apply the
formulae (83)-(84) for the Hermite polynomials and combining the above
formulae we have the known result for the photon distribution function
in squeezed and correlated state
\beq
P_{n}=P_{0}\frac{(\tanh r)^{n}}{n!2^{n}}|H_{n}(e^{\frac{-i\theta}{2}}
\sqrt {\tanh r}[\frac{<q>-i<p>}{2} +e^{i\theta }\coth r \frac{<q>+i<p>}{2}])
|^{2}.
\eeq
For squeezed vacuum state $<p>=<q>=0$  the argument of Hermite polynomial
is equal to zero and using the formula (78) for $\theta =0$ we obtain in this
case the probability to have even numbers of photons
\beq
P_{2n}^{vac}=\frac{1}{\cosh r}(\frac{\tanh r}{2})^{n}\frac{2n!}{(n!)^{2}}.
\eeq
The probability to have odd number of photons is equal to zero due to
the relation (85).
The mean values of photons and the dispersion of photon number are described
by the formulae (96) and (97) respectively in which the matrix $\mu $ has
the form
\beq
\mu =\cosh ^{2}r\left( \begin{array}{clcr}
1 & {-e^{i\theta }\tanh r}\\
{-e^{-i\theta }\tanh r} & 1\end{array} \right)
\eeq
and the vector ${\bf y}$ has the components given by the formula (24).The
number $\nu $ is given by the formula (99) and it is equal
\beq
\nu =-\sinh ^{2}r.
\eeq
For the coherent state $r=0$ and the matrix $\mu $ takes the form
\beq
\mu =\left( \begin{array}{clcr}
1 & 0\\
0 & 1\end{array} \right).
\eeq
The number $\nu $ is equal to zero since the matrix $R=0$.So,for coherent
states the general formula for photon mean value (96) gives
\beq
<n>=\frac{1}{2}(<p>^{2}+<q>^{2})
\eeq
since in this case $(R{\bf y})_{1}=\frac{1}{\sqrt 2}(<q>+i<p>)$.The
dispersion given by the formula (97) is equal to the photon number mean value
that corresponds to the Poisson distribution function.If there is no
correlation $(\theta =0)$ the formula (96) gives the mean value of
photons for the squeezed light
\beq
<n>=\sinh ^{2}r +\frac{<p>^{2}+<q>^{2}}{2}.
\eeq
If there is correlation,i.e.the correlation coefficient $k$ (107) is not
equal to zero $(\theta \neq 0)$ the formula (96) gives the same
mean value of
photons (117)for squeezed and correlated light.One can show that the
formula (96) yields for mean value of photons in
generic case the following expression
\beq
<n>=\frac {T-1}{2}+\frac {1}{2} <{\bf Q}><{\bf Q}>.
\eeq

The dispersion $\sigma _{n}$ is described by the formula (97) with the
matrix $\mu $ (113),the number $\nu $ (114),and it can be calculated
due to the following relation
\beq
\sigma_{n}=\frac {T^{2}}{2}-d-\frac {1}{4}+<{\bf Q}>m<{\bf Q}>
\eeq
which is valid in generic case.

\section{Photon distribution for black body radiation}

For all the pure states which are described by the gaussian Wigner function
the matrix $R$ is diagonal since the determinant of the dispersion matrix
$d=\frac{1}{4}$ for these states.The parameter which is used to distinguish
the mixed states is the following
\beq
\eta=Tr~\hat \rho ^{2}.
\eeq
For one-dimensional system with gaussian Wigner function the parameter
is determined by the determinant of the dispersion matrix $m$.It is
\beq
\eta=\frac{1}{2\pi }\int W^{2}(p,q)dpdq=\frac{1}{2\sqrt d}.
\eeq
For $d=\frac{1}{4}$ the parameter $\eta =1$ that corresponds to the pure
states.The obvious relation
\beq
Tr~\hat \rho ^{2}\leq Tr~\hat \rho =1,~~~~d\geq \frac{1}{4}
\eeq
gives the uncertainty relation with the correlation coefficient $k$
\beq
\sigma _{pp}\sigma _{qq}\geq \frac{1}{4}(\frac{1}{1-k^{2}}),~~~\hbar =1,
\eeq
which is Schrodinger uncertainty relation [19].So,when the parameter
$\eta \rightarrow 1$ the matrix $R$ takes the diagonal form and the
inequality (123) transforms to the equality
\beq
\sigma _{pp}\sigma _{qq}=\frac{1}{4}(\frac{1}{1-k^{2}}).
\eeq
For all the pure states with gaussian Wigner function the $Q$-photon
function takes the factorized form since the matrix $R$ is the diagonal
one.\\
For the thermal equilibrium state the dispersion matrix is
\beq
m_{t}=\frac{1}{2}\coth \frac{\beta }{2}\left( \begin{array}{clcr}
1&0\\
0&1\end{array} \right).
\eeq
 The corresponding matrix $R_{t}$ has the form
\beq
R_{t}=-e^{-\beta}\left( \begin{array}{clcr}
0&1\\
1&0\end{array} \right)
\eeq
and according to (24) the arguments of Hermite polynomials in the general
expression for the photon probabilities are in this thermal equilibrium
state
\beq
y_{1}=y_{2}^{*}=-{\sqrt {2}}(<q >-i<p>)/(1-\coth \frac {\beta }{2}).
\eeq
Combining the Eq.(127) with the formula (90) for the Hermite polynomial
with the matrix $R$ proportional to the Pauli matrix $\sigma _x$ we
have in this case
\beq
H_{nn}^{\{R\}}(y_{1},y_{1}^{*})=(-1)^{n}n!e^{-\beta n}
L_{n}(e^{-\frac {\beta }{2}}(\cosh \beta -1)(<p>^{2}+<q>^{2})).
\eeq
So the formula (28) gives the photon distribution for the limiting case
of the field thermal state
\beq
P_{n}=(-1)^{n}2(1+\coth \frac {\beta }{2})^{-1}exp[-\frac {<p>^{2}+<q>^{2}}
{1+\coth \frac {\beta }{2}}-\beta n]L_{n}(e^{-\frac {\beta }{2}}(\cosh \beta
-1)
(<p>^{2}+<q>^{2})).
\eeq
In the termal equilibrium state the matrix $\mu $ (98) is proportional
to the identity matrix
\beq
\mu = \frac {1}{1-e^{-\beta }}\bf 1
\eeq
and the number $\nu $ (99) is
\beq
\nu =\frac {1}{1-e^{\beta }}.
\eeq
So,from the general expression (96) we have the mean value of photon number
in the thermal equilibrium state
\beq
<n>=\frac {1}{e^{\beta }-1}+\frac {<q>^{2}+<p>^{2}}{2}
\eeq
where the first term is the Planck distribution and the correction to it is
related to the shift of the frame of reference in the phase space of the
electromagnetic field oscillator.Obviously this correction does not depend
on the temperature.The dispersion $\sigma _{n}$ may be obtained using
the above numbers and the formula(97).We have
\beq
\sigma _{n}=\frac {e^{\beta }}{(e^{\beta }-1)^{2}}+
\frac {e^{\beta }+1}{(e^{\beta }-1)}(\frac {<p>^{2}+<q>^{2}}{2}).
\eeq

\section{Squeezed and correlated vacuum state}

We will consider the last limiting case of squeezed and correlated vacuum
state corresponding to the most general Wigner function for which the
quadrature dispersion matrix m has 4 nonzero matrix elements and its
determinant is not equal to $\frac {1}{4}$.But the quadrature means
$<p>$ and $<q>$ are taken to be zero.This state corresponds to the factorized
form of the density operator (50) with the complex parameter $<z>=0$.In this
case due to the formula (28) the photon distribution function is given by the
expression
\beq
P_n=P_{0}\frac {H_{nn}^{\{R\}}(0,0)}{n!}
\eeq
where the arguments of Hermite polynomial $y_1$ and $y_2$ are equal to zero
due to the relation (24) in which $<z>=0$.The probability to have no photons
in the mixed sqeezed vacuum state is given by the expression (26) which takes
in this case the form
\beq
P_0=(d+\frac {1}{2} T+\frac {1}{4})^{-\frac{1}{2}}.
\eeq
So,this probabability depends on two invariant parameters
only.One parameter is the determinant of the quadrature dispersion matrix
and another one is the trace of this matrix.The matrix $R$ has the nondiagonal
matrix elements (23) depending on the same invariant parameters and the
diagonal
matrix element (23) depending on the third parameter-the correlation
coefficient of the two quadratures which is proportional to $\sigma _{pq}$.
For the mixed state this term is not the function of the parameters
$\sigma_{pp}$ and $\sigma _{qq}$ only as it takes place for any pure Gaussian
state.
But we will see below that the distribution function depends only on two
parameters.

There exists the relation [9] of two Hermite polynomials of two variables
$H_{nm}^{\{R\}}(x,y)$ and $H_{nm}^{\{R_{1}\}}(x_{1},y_{1})$ for which
the three matrix elements of the symmetric matrix $R$
namely $R_{11}$,$R_{22}$ and
$R_{12}$ are arbitrary numbers  but the matrix $R_{1}$ has the matrix elements
$(R_{1})_{11}=(R_{1})_{22}=1$ and
\beq
(R_{1})_{12}=\frac {R_{12}}{\sqrt {R_{11}R_{22}}}.
\eeq
This relation has
the form
\beq
H_{nm}^{\{R\}}(x,y)=R_{11}^{n/2}R_{22}^{m/2}H_{nm}^{\{R_{1}\}}(x_{1},y_{1})
\eeq
where
\beq
x_1=x\sqrt {R_{11}},~~~~~~y_1=y\sqrt {R_{22}}.
\eeq
Due to the formula (76) for the matrix $R_1$ the corresponding
Hermite polynomial $H_{nn}^{\{R_{1}\}}(0,0)$ is expressed in terms of
Legendre polynomials $P_{n}(x)$.
So,we have for the case of generic matrix $R$ the expression
\beq
H_{nn}^{\{R\}}(0,0)=n!(-D)^{n/2}P_{n}(\frac {-R_{12}}{\sqrt {-D}})
\eeq
where $D=R_{11}R_{22}-R_{12}^2$.Using this relation we can obtain
from the general formula (28) the photon distribution function for the
generic mixed vacuum state
\beq
P_{n}=P_{0}(\frac {2d+\frac {1}{2}-T}{2d+\frac {1}{2}+T})^{n/2}
P_{n}(\frac {2d-\frac {1}{2}}{\sqrt {(2d+\frac {1}{2})^{2}-T^{2}}}).
\eeq
Here the parameters $T$ and $d$ are expressed in terms of the
dispersion matrix $m$ (see formulae (4),(5)).This expression for the field
distribution has been obtained in other notations in Ref.[4],[9].
So,for pure state (the temperature $\beta ^{-1}$ is equal to zero) we have
the photon distribution function taking nonzero values only for even
number of photons since the Legendre polynomials $P_{2k+1}(0)$ are equal
to zero.That corresponds to the squeezed and correlated vacuum state which
is the superposition of even photon number states.The temperature changes
the photon distribution function and there are nonzero probabilities
to have odd number of photons for mixed squeezed and correlated
vacuum state.Thus for the probability to have one photon the formula
(140) gives the value
\beq
P_1=\frac {2d-\frac {1}{2}}
{(2d+\frac {1}{2}+T)\sqrt {\frac {T}{2}+\frac {1}{4}+d}}.
\eeq
The photon distribution function for squeezed and correlated vacuum
state depends on one parameter $T$ only but for the mixed vacuum state it
depends on two invariant parameters $T$ and $d$.
If the parameter $\eta $ (121) slightly differs from unity we have
\beq
1-\eta =\gamma ,~~~~~~~~~0<\gamma<<1.
\eeq
We could consider the corrections to the photon distribution functin of
pure states connected with the influence of the mixing parameter $\gamma $.
Thus the probability to have one photon is given by approximate formula
\beq
P_1=\frac {\gamma }{(1+T)\sqrt {\frac {1}{2}(1+T)}}.
\eeq
All the probabilities to have odd number of photons are proportional
to the mixing parameter $\gamma $.Thus measuring these probabilities
we could measure the degree of the mixing of the photon system state.
If the photon state is obtained by parametric excitation of an initial
thermal state it is equivalent to measuring of the temperature
of the initial state due to the relations (32),(33) and (39).

\section{Photon distribution for generic Gaussian state}

In this section we will concider the photon distribution function (28)
expressed in terms of the series of the products either of two Hermite
polynomials or two Laguerre polynomials.

The formula (67) gives the expression of the Hermite polynomial of two
variables with equal indecies in terms of the products of two
Hermite polynomials of one variable.Using this formula one obtains
from the general expression (28) the following photon distribution
function
\beq
P_{n}=P_{0}n!\sum_{k=0}^{n}(\frac {R_{11}R_{22}}{4})^{n/2}  (-\frac {2R_{12}}{
\sqrt {R_{11}R_{22}}})^{k}(n-k)!^{-2}k!^{-1}
|H_{n-k}(\frac {R_{11}y_{1}+R_{12}y_{2}}{\sqrt {2R_{11}}})|^{2}.
\eeq
Here the factor $P_{0}$ is given by the formula (26),the matrix elements of
the matrix $R$ are given by the formulae (22),(23) and the numbers $y_{1}$,
$y_{2}$ are given by the formula (24).If the state is the pure squeezed and
correlated state the determinant of the quadrature dispersion matrix is
equal to $\frac{1}{4}$ and the matrix element $R_{12}=0$ due to the formula
(23).In this case only the first term in the series (144) is not equal to zero
and this term gives the expression (111) for the photon distribution function
of the pure and correlated state obtained by another method in section 8.The
expression (144) is convinient to discuss the connection of the uncertainty
relations with the quantum distribution functions.In fact the formula
for the Gaussian Wigner function (6) formally coincides with the classical
Gaussian density of probability in the particle phase space.Then we have
to answer the question in what aspects these formulae must be considered
as essentially different ones.By intuition it is clear that the uncertainty
relation must distinguish the classical and quantum Gaussian densities
of probabilities having absolutely the same form (6).And in fact in addition
to the usual restriction for the dispersion matrix for the Gaussian
classical distribution function in the phase space which is the condition
of nonnegativity of this dispersion matrix the quantum mechanics demands
the inequality for the determinant of this matrix (122).This inequality
is the Schrodinger uncertainty relation.The formula (144) permits
to relate this inequality with physically obvious condition that
probability to find $n$ photons must be nonnegative because the determinant
of the dispersion matrix is the parameter on which depends the photon
distribution function.We see that all the terms in the expression (144)
are obviously positive ones except the term containing
the number $-2R_{12}$ which may change the sign for odd powers $k$ if
it is not positive itself.It means that for
natural condition of positiveness of photon distribution function it
is nesessary to have inequality
\beq
R_{12}<0
\eeq
But this inequlity is equivalent to the inequality for the dispersion
matrix determinant (122) which implies the uncertainty relation.Thus
existence of the connection of the Wigner function (6) with the photon
distribution funtion (144) is consistent only if the uncertainty relation
holds.So we clarify the mechanism how uncertainty relation in
phase space of electromagnetic field oscillator
influences the form of photon distribution function.

Another expression for the photon distribution function may be derived
from the general expression (28) if one uses the formula (75) for the Hermite
polynomial of two variables in terms of Laguerre polynomials.Then we have
\beq
P_{n}=P_{0}(-1)^{n}\sum_{s=0}^{n} (R_{12}-\sqrt {R_{11}R_{22}})^{s}
(R_{12}+\sqrt {R_{11}R_{22}})^{n-s}L_{s}^{-\frac {1}{2}}(x_{1})L_{n-s}^
{-\frac{1}{2}}(x_{2})
\eeq
where the factor $P_{0}$ is given by the formula (26),the matrix elements
of the matrix $R$ are given by the formulae (22),(23) and the numbers
$x_1$,$x_2$ are given by the relations (71) and (72) in which one
has to replace $y_1$ by $R_{11}y_{1}+R_{12}y_{2}$ and $y_{2}$ by
$R_{12}y_{1}+R_{22}y_{2}$.In this form the photon distribution
function has been obtained in Ref.[4].
As we have shown the series in this expression may be summed and the sum is
equal to Hermite polynomial of two variables.

The distribution function (28) demonstrates wavy behaviour for the mixed light
similar to the wavy behaviour of the distribution function found
in Ref.[20],[21] for pure squeezed and correlated states of the light.

In conclusion we could point out that
the results obtained for the one-mode case may be extended for the polymode
case since the structure of general formula for the photon distribution
function
of polymode mixed Gaussian light is described by Hermite polynomial
of $2N$ variables [9].

\section{Acknoledgements}
One of us (V.I.M.)thanks INFN and University of Napoli "Federico II"
for the hospitality.

\begin {center}
{\LARGE\bf References}\\
\end {center}

[1] G.S.Agarwal,J.Mod.Opt.{\bf {34}},909(1987).

[2] G.S.Agarwal and G.Adam,Phys.Rev.A {\bf {38}},750(1988).

[3] G.S.Agarwal and G.Adam,Phys.Rev.A {\bf {39}},6259(1989).

[4] S.Chaturvedi and V.Srinivasan,Phys.Rev.A {\bf {40}},6095(1989).

[5] P.Marian and T.A.Marian,Phys.Rev.A {\bf 47},4474(1993).

[6] A.Vourdas,Phys.Rev.A {\bf 34},3466(1986);Phys.Rev.A {\bf 36},5866(1987).

[7] H.Fearn and M.J.Collett,J.Mod.Opt.{\bf 35},553(1988).

[8] M.S.Kim,F.A.M.Oliveira and P.L.Knight,Phys.Rev.A {\bf 40},2494(1989).

[9] V.V.Dodonov,V.I.Man'ko and V.V.Semjonov,
Nuovo Cimento B {\bf {83}},145(1984).

[10] G.Adam,Phys.Lett.A {\bf 171},66(1992).

[11] R.J.Glauber,Phys.Rev.Letters {\bf {10}},84(1963).

[12] D.Stoler,Phys.Rev.D {\bf {1}},3217(1970).

[13] P.P.Bertrand,K.Moy and E.A.Mishkin,Phys.Rev.D {\bf {4}},1909(1971).

[14] H.P.Yuen,Phys.Rev.A {\bf {13}},2226(1976).

[15] J.N.Hollenhorst,Phys.Rev.D {\bf {19}},1669(1979).

[16] V.V.Dodonov,E.A.Kurmyshev and V.I.Man'ko,Phys.Lett.A {\bf {79}},150(1980).

[17] V.V.Dodonov and V.I.Man'ko "Invariants and evolution of
nonstationary quantum systems",Proceedings of Lebedev Physical institute
{\bf 183},ed.by M.A.Markov,Nova Science Publishers,
Commack,N.Y.(1989).

[18] Bateman Manuscript Project:Higher Transcendental Functions,
edited by A.Erdely (McGraw-Hill,New York,N.Y.(1953).

[19] E.Schrodinger,Ber.Kgl.Akad.Wiss.Berlin,296(1930).

[20] W.Schleich and J.A.Wheeler,J.Opt.Soc.Am.B {\bf {4}},1715(1987).

[21] V.V.Dodonov,A.B.Klimov and V.I.Man'ko,Phys.Lett.A {\bf {134}},211(1989).

\end{document}